*Article*

# The normalization of citation counts based on classification systems

**Lutz Bornmann** [1,*,†], **Werner Marx** [2,†], **Andreas Barth** [3,†]

[1] Division for Science and Innovation Studies, Administrative Headquarters of the Max Planck Society, Hofgartenstr. 8, 80539 Munich (Germany), E-Mail: bornmann@gv.mpg.de

[2] Max Planck Institute for Solid State Research, Heisenbergstrasse 1, D-70569 Stuttgart (Germany), E-Mail: w.marx@fkf.mpg.de

[3] FIZ Karlsruhe, Hermann-von-Helmholtz-Platz 1, D-76344 Eggenstein-Leopoldshafen (Germany), E-Mail: andreas.barth@fiz-karlsruhe.de

[†] These authors contributed equally to this work

[*] Author to whom correspondence should be addressed; E-Mail: bornmann@gv.mpg.de

**Abstract:** If we want to assess whether the paper in question has had a particularly high or low citation impact compared to other papers, the standard practice in bibliometrics is to normalize citations in respect of the subject category and publication year. A number of proposals for an improved procedure in the normalization of citation impact have been put forward in recent years. Against the background of these proposals this study describes an ideal solution for the normalization of citation impact: in a first step, the reference set for the publication in question is collated by means of a classification scheme, where every publication is associated with a single principal research field or subfield entry (e. g. via Chemical Abstracts sections) and a publication year. In a second step, percentiles of citation counts are calculated for this set and used to assign the normalized citation impact score to the publications (and also to the publication in question).

# 1. Introduction

If we wish to assess whether a paper has had a particularly high or particularly low citation impact compared to other papers, the standard practice in bibliometrics is to normalize citation counts which are field-specific. "Each field has its own publication, citation and authorship practices, making it difficult to ensure the fairness of between-field comparisons. In some fields, researchers tend to publish a lot, often as part of larger collaborative teams. In other fields, collaboration takes place only at relatively small scales, usually involving no more than a few researchers, and the average publication output per researcher is significantly lower. Also, in some fields, publications tend to have long reference lists, with many references to recent work. In other fields, reference lists may be much shorter, or they may point mainly to older work. In the latter fields, publications on average will receive only a relatively small number of citations, while in the former fields, the average number of citations per publication will be much larger" [1]. The citation impact can only be used for cross-field comparisons after a field-specific normalization of the citation impact of papers has been undertaken. In bibliometrics, normalizing procedures use statistical methods to calculate citation impact values which are comparable across different fields and times.

## 2. The use of reference sets

When normalized citation indicators are generated, two standards are used as reference sets: "the average citation rate of the journal and the average citation rate of the field. In the first case, the citation count of a paper is compared with the average citation rate for the particular journal in the particular year. In the second case, the citation count of a paper is compared with the average citation rate for the particular field or subfield for the particular year" [2, p. 172]. These two standards are then used to calculate relative citation indices, which are called the Relative Citation Rate [3]. With this standard one has to take into account that most bibliometric studies are based on journals: either they are based on individual journals, or journal sets are used, where individual journals are combined to form a field-specific set. "Primary journals in science are generally agreed to contain coherent sets of papers both in topics and in professional standards. This coherence stems from the fact that many journals are nowadays specialized in quite narrow sub-disciplines and their 'gatekeepers' (i.e. the editors and referees) controlling the journal are members of an 'invisible college' sharing their views on questions like relevance, validity or quality" [4, p. 314].

Even though both standards (normalization based on individual journals or on field-specific journal sets) have already been used in a variety of ways in bibliometrics, there are a number of arguments in favour of a preference for the use of research fields in the normalization rather than the use of individual journals:

1) Indicators represent incentives for academics to shape their publication behaviour in a particular way [5]. Academics are therefore guided by the indicators which are used in research evaluations. The normalization based on individual journals rewards publications in journals of little reputation: in these journals it is easier for individual publications to be above the reference value [6]. The use of an indicator which is normalized on the basis of individual journals therefore encourages academics to publish in journals of lesser reputation.

2) In general, reference values should be used to take account of (or to disregard) factors in the citation analysis which may have an impact on citations but are not related to research quality. The year of publication affects the citation impact of a publication, for example, although the year of publication has no bearing on quality. We can assume that a publication from 2000 is not of a higher quality per se than a publication from 2005 - even if the older publication is usually cited more often than the more recent one. We also know (see above) that the research field has an influence on citations. The different citation rates between the research fields do not reflect differences in quality between the papers in the research fields, however. Whereas mean citation impact values for different subject categories reflect only the different citation behaviours within different research fields, the values for different individual journals reflect not only the different behaviours, but also the different journal qualities. We know that certain journals publish (on average) higher quality papers than other journals [7]. Thus, the citation impact score for a journal is also quality driven, but not the citation impact score for a research field. This feature leads to the fact that results on the basis of standards which are based on individual journals are not meaningful without accompanying indicators.

3) Indicators on the basis of a normalization based on individual journals must therefore always be accompanied by indicators which provide information on the quality of the journals where the research under evaluation has been published. The normalized score on its own is not meaningful: if two institutions A and B have the same above-average score, it is not clear whether the score is based on normalization to journals with high or low citation impact. Institution A, which was normalized to a high citation impact would have published in reputational journals and at the same time achieved more citations. This institution would therefore be successful in two respects. Institution B, which was normalized to a low citation impact, would have published in unimportant journals and exceeded only this low standard. Institution B has in fact a worse position (in two respects), a fact which is not expressed by the normalized score [8]. Only the quality of the journals enables an assessment to be made as to whether an institution has truly achieved a high impact with its publications when it has a comparably low normalized citation impact (because it has published mainly in reputational journals with a high citation impact), or whether it has truly achieved a low impact (because it has published mainly in journals of little reputation with a low citation impact).

4) In bibliometric evaluations the mean normalized citation impact (of institutions, for example) is often shown as a function of the individual years of publication. Since individual journals usually enter into the calculation of normalized impact scores with significantly smaller publication sets than do journal sets, this leads to the normalized scores based on individual journals exhibiting greater variations over the publication years than the normalized scores based on the publications in a specific research field. The variations often make it almost impossible to recognize a true trend over the publication years for normalized scores based on individual journals.

Given these problems, other publications have already recommended that the field-normalization should be given preference over normalization based on single journals: Aksnes [2], for example, writes that "the field average should be considered as a more adequate or fair baseline [than the Relative Citation Rate], a conclusion that is also supported by other studies" (p. 175). The Council of Canadian Academies [9] recommends that "for an assessment of the scientific impact of research in a field at the national level, indicators based on relative, field-normalized citations (e.g., average relative citations) offer the best available metrics. At this level of aggregation, when appropriately normalized by field and based on a sufficiently long citation window, these measures provide a defensible and informative assessment of the impacts of past research" (p. xv).

The preference for the use of a research field instead of an individual journal in bibliometrics does not mean that the end of the discussion on normalized indicators has now been reached, however. Recent proposals for improving the calculation of normalized impact scores refer primarily to (1) the use of better alternatives for journal sets and (2) the avoidance of the arithmetic average when calculating reference scores.

## 3. Determination of research fields

In most studies the determination of research fields is based on a classification of journals into subject categories developed by Thomson Reuters (Web of Science) or Elsevier (Scopus). "The Centre for Science and Technology Studies (CWTS) at Leiden University, the Information Science and Scientometrics Research Unit (ISSRU) at Budapest, and Thomson Scientific [now Thomson Reuters] itself use in their bibliometric analyses reference standards based on journal classification schemes" [10]. Each journal is classified as a whole either to one or to several subject categories. The limitations of journal classification schemes become obvious in the case of multidisciplinary journals such as Nature or Science and highly specialized fields of research. Papers that appear in multidisciplinary journals cannot be assigned exclusively to one field, and for highly specialized research fields no adequate reference values exist. To overcome the limitations of journal classification schemes, Bornmann, *et al.* [11] and Neuhaus and Daniel [10] proposed an alternative possibility for the compilation of comparable sets of publications (the reference standard) for the papers in question. Their normalization is based on a publication-specific classification where each publication is associated with at least one single principal field or subfield entry, highlighting the most important aspect of the individual publication [12].

The databases offered by Chemical Abstracts Service (CAS), a division of the American Chemical Society (ACS), are the most comprehensive databases of publicly disclosed research in chemistry and related sciences. The CAS literature database (CAplus) includes both papers and patents published since around 1900. This database not only covers publications in the classical fields of chemistry, but also in many other natural science disciplines like materials science, physics and biology. CAS has defined a three-level classification scheme to categorize chemistry-related publications into five broad headings of chemical research (section headings) which are divided in 80 different subject areas called Chemical Abstracts sections. (see Table 1). Each of the 80 sections is further divided into a varying number of sub-sections. Each individual paper is assigned to only one section or subsection according to the main subject field and interest. If the subject matter is appropriate to other sections, cross-references are provided. Detailed descriptions of all sections can be found on the CAS webpage and in Chemical Abstracts Service [13]. This classification is applied to all publications of the CAplus literature and patent database (see https://www.cas.org/content/ca-sections):

- "Each CA section covers only one broad area of scientific inquiry
- Each abstract in CA appears in only one section
- Abstracts are assigned to a section according to the novelty of the process or substance that is being reported in the literature
- If abstract information pertains to a section(s) in addition to the one assigned, a cross-reference is established"

**Table 1.** Summary of CA Section Headings.
For Organic Chemistry the individual sections are listed for illustration.

| Section Heading | Number of Sections |
|---|:---:|
| **BIOCHEMISTRY (BIO/SC)** | 20 |
| **ORGANIC (ORG/SC)** | 14 |
| 21. **General Organic Chemistry** | |
| 22. **Physical Organic Chemistry** | |
| 23. **Aliphatic Compounds** | |
| 24. **Alicyclic Compounds** | |
| 25. **Benzene, Its Derivatives, and Condensed Benzenoid Compounds** | |
| 26. **Biomolecules and Their Synthetic Analogs** | |
| 27. **Heterocyclic Compounds (One Hetero Atom)** | |
| 28. **Heterocyclic Compounds (More Than One Hetero Atom)** | |
| 29. **Organometallic and Organometalloidal Compounds** | |
| 30. **Terpenes and Terpenoids** | |
| 31. **Alkaloids** | |
| 32. **Steroids** | |
| 33. **Carbohydrates** | |
| 34. **Amino Acids, Peptides, and Proteins** | |
| **MACROMOLECULAR (MAC/SC)** | 12 |
| **APPLIED (APP/SC)** | 18 |
| **PHYSICAL, INORGANIC, AND ANALYTICAL (PIA/SC)** | 16 |

The number of papers per section and year varies largely. E.g., in 2010 the average number of papers per section is 11869, with 73273 (section 1) as the highest and 240 (section 32) as the lowest number of papers. From a statistical point of view, this is widely sufficient for a reliable normalization. If sections are too large, it is questionable whether sections are sufficiently homogeneous in terms of citation practices (this has to be investigated further). It could be that a section covers subareas of chemical research with different citation practices. In any case, the sub-sections rather the complete sections can be consulted for normalization, provided that the number of papers per year meets statistical requirements [14].

An advantage of the sections of Chemical Abstracts for bibliometric analyses is that indexers assign the relevant sections to the papers intellectually. This classification is not affected by what is called the "indexer effect": According to Braam and Bruil [15], the classification of papers into 80 sections in Chemical Abstracts is in accordance with author preferences for 80% of all papers. The sections of Chemical Abstracts thus seem to provide a promising basis for the description and comparison of publications and impact profiles of journals. Hence, for evaluation studies in the field of chemistry and related fields [14,16], comparable papers can be compiled to reference sets using a specific CA section or sub-section which covers the content of this specific publication set to a large extent. In contrast to the classification of journals in journal sets, this procedure also assigns papers from multidisciplinary and wide-scope journals to a specific field.

In addition to Chemical Abstracts there are a number of other specialist databases which categorize publications on a paper-by-paper basis in terms of research fields (e. g. Medline or Scitation). In the field of Mathematics a common Mathematical Subject Classification (MSC) has been developed by the two providers of large mathematical literature databases: Zentralblatt Math from FIZ Karlsruhe and MathSciNet from the American Mathematical Society (http://msc2010.org). After a revision in 2010 both providers apply the same classification codes to all documents in their databases. On the top level there are 63 subject headings with a considerable number of specific classification codes. Each document includes at least one MSC. Since the MSC is a rather detailed set of classifications which are systematically applied it should be well suited to a bibliometric analysis of papers with respect to their research fields.

Scopus manually adds index terms from different specialist areas for the majority of the titles included in Scopus. These index terms are adopted from thesauri which the database operator Elsevier itself owns or licenses (e. g. Medline). The terms are added to the publication records in order to improve retrieval from a field-specific search for publications. In addition to MeSH terms (Medline) for the fields of life sciences and health sciences the EI thesaurus, for example, is used for the fields of engineering, technology and the physical sciences.

In some fields, it may be difficult to find one complete paper classification scheme. In these fields, several schemes must be indicated. For example, in mathematics a major scheme exists, but in economics it is hard to find an appropriate scheme. Moreover, some fields (e.g. information sciences) do not have a classification scheme that is accepted by everyone.

## 3. Percentiles of citation counts

Two significant disadvantages are inherent in the calculation of the Relative Citation Rate [7]: (i) As a rule, the distribution of citations over publication sets is skewed to the right. The arithmetic mean value calculated for a reference set is therefore determined by a few highly cited papers. The arithmetic mean as a measure of central tendency is not suitable for skewed data. This is the only reason why, for example, in the Leiden Ranking 2011/2012 the University of Göttingen occupies position 2 in a ranking by citation impact; the relevant mean score for this university "turns out to have been strongly influenced by a single extremely highly cited publication" [17, p. 2425]. (ii) The quotient permits merely a statement about whether a publication is cited more or less than the average in the reference set. Other attributes which could describe the citation impact of a publication as excellent or outstanding are based on (arbitrary) rules of thumb with no relationship to statistical citation distributions [18].

Using percentiles (or percentile rank classes) to normalize citation impact can give better comparisons of the impact of publications than normalization using the arithmetic mean [19-22]. The percentile provides information about the citation impact the publication in question has had compared to other publications. A percentile is a value below which a certain proportion of observations fall: the higher the percentile for a publication, the more citations it has received compared to publications in the same research field and publication year. The percentile for a publication is determined using the distribution of the percentile ranks over all publications: for example, a value of 90 means that the publication in question is among the 10% most cited publications; the other 90% of the publications have achieved less impact. A value of 50 represents the median and thus an average citation impact compared to the other publications (from the same research field and publication year).

For a publication set under study, each publication in the set must be normalized using its specific reference set with publications from the same field and publication year. In other words, each publication receives its specific percentile which is calculated based on its specific reference set. However, this normalization using percentiles is sensitive to the coverage of a bibliographic database. For instance, the more local or national journals (typically of low impact) a database contains, the easier it becomes for a publication in an international journal (typically of a higher impact) to have a high percentile rank [23]. There simply are more lowly cited publications and therefore a publication that is sufficiently highly cited will end up in a higher percentile. What is even more problematic is that there can be differences between fields in database coverage. In some fields there may be many local or national journals covered by a database, making it relatively easy to end up in a high percentile, whereas in other fields there may be only a few local or national journals, making it more difficult to end up in a high percentile. In some fields a database such as Web of Science also covers popular magazines (e.g., *Forbes* and *Fortune in business*). These magazines, which can hardly be considered scientific, receive few citations and therefore it becomes relatively easy for other publications in these fields to end up in high percentiles. These issues should be considered if a certain database is selected for a specific research evaluation study.

## 4. Conclusions

Given the new possibilities for calculating reference values and the strengths and weaknesses of existing standard indicators several recent research papers have proposed alternative solutions for the normalization of citation counts. It is one object of current research to compare the different methods empirically and to find the "best" field-normalizing method. For example, Leydesdorff, *et al.* [24] compare normalization by counting citations in proportion to the length of the reference list ($1/N$ of references) with rescaling by dividing citation scores by the arithmetic mean of the citation rate [25]. The former normalization method uses the citing papers as the reference sets across fields and journals, and then attributes citations fractionally from this perspective. In the latter normalization method proposed by Radicchi, *et al.* [26], the normalized (field-specific) citation count is $c_f = c / c_0$, in which c is the raw citation count and $c_0$ is the average number of citations per unit (article, journal, etc.) for this field – or more generally – this subset. The results of Leydesdorff, Radicchi, Bornmann, Castellano and de Nooy [24] show, for example, that rescaling outperforms fractional counting of citations.

Our approach is based on a high-quality classification system which is intellectually and systematically applied to all publications in a given database. The normalization of the citation impact can be carried out in two steps: in a first step, the reference set for the publication in question is collated by means of a classification scheme, where every publication is associated with a single principal field or subfield entry, e. g. via CA sections. In a second step, percentiles are calculated for this set, and are then used to assign a normalized citation impact score to the publication in question.

This approach offers a simple operational solution for the normalization of the citation impact [7]. It provides a significant improvement with respect to both existing solutions (journal or field based) as well as to other approaches currently under investigation. The major advantages are the application of a systematic high-quality classification system, the simplicity of the procedure, and most importantly the balance or fairness of the resulting citation counts.

## Conflict of Interest

The authors declare no conflict of interest.

## References

1.  Waltman, L.; van Eck, N.J. A systematic empirical comparison of different approaches for normalizing citation impact indicators. http://arxiv.org/abs/1301.4941 (February 6),
2.  Aksnes, D.W., Citation rates and perceptions of scientific contribution. *Journal of the American Society for Information Science and Technology* **2006**, *57*, 169-185.
3.  Schubert, A.; Braun, T., Relative indicators and relational charts for comparative assessment of publication output and citation impact. *Scientometrics* **1986**, *9*, 281-291.
4.  Schubert, A.; Braun, T., Cross-field normalization of scientometric indicators. *Scientometrics* **1996**, *36*, 311-324.
5.  Bornmann, L., Mimicry in science? *Scientometrics* **2010**, *86*, 173-177.
6.  Vinkler, P., The case of scientometricians with the "absolute relative" impact indicator. *J. Informetr.* **2012**, *6*, 254-264.
7.  Bornmann, L.; Mutz, R.; Marx, W.; Schier, H.; Daniel, H.-D., A multilevel modelling approach to investigating the predictive validity of editorial decisions: Do the editors of a high-profile journal select manuscripts that are highly cited after publication? *Journal of the Royal Statistical Society - Series A (Statistics in Society)* **2011**, *174*, 857-879.
8.  van Raan, A.F.J., Measurement of central aspects of scientific research: Performance, interdisciplinarity, structure. *Measurement* **2005**, *3*, 1-19.
9.  Council of Canadian Academies *Informing research choices: Indicators and judgment: The expert panel on science performance and research funding.* ; Council of Canadian Academies: Ottawa, Canada, 2012.
10. Neuhaus, C.; Daniel, H.-D., A new reference standard for citation analysis in chemistry and related fields based on the sections of chemical abstracts. *Scientometrics* **2009**, *78*, 219-229.
11. Bornmann, L.; Mutz, R.; Neuhaus, C.; Daniel, H.-D., Use of citation counts for research evaluation: Standards of good practice for analyzing bibliometric data and presenting and interpreting results. *Ethics in Science and Environmental Politics* **2008**, *8*, 93-102.
12. van Leeuwen, T.N.; Calero Medina, C., Redefining the field of economics: Improving field normalization for the application of bibliometric techniques in the field of economics. *Res. Evaluat.* **2012**, *21*, 61-70.
13. Chemical Abstracts Service *Subject coverage and arrangement of abstracts by sections in chemical abstracts*; Chemical Abstracts Service (CAS): Columbus, OH, USA, 1997.
14. Bornmann, L.; Schier, H.; Marx, W.; Daniel, H.-D., Is interactive open access publishing able to identify high-impact submissions? A study on the predictive validity of *atmospheric chemistry and physics* by using percentile rank classes. *Journal of the American Society for Information Science and Technology* **2011**, *62*, 61-71.
15. Braam, R.R.; Bruil, J., Quality of indexing information: Authors views on indexing of their articles in chemical abstracts online ca-file. *J. Inf. Sci.* **1992**, *18*, 399-408.
16. Bornmann, L.; Daniel, H.-D., Selecting manuscripts for a high impact journal through peer review: A citation analysis of communications that were accepted by *angewandte chemie* *international edition*, or rejected but published elsewhere. *Journal of the American Society for Information Science and Technology* **2008**, *59*, 1841-1852.

17.    Waltman, L.; Calero-Medina, C.; Kosten, J.; Noyons, E.C.M.; Tijssen, R.J.W.; van Eck, N.J.; van Leeuwen, T.N.; van Raan, A.F.J.; Visser, M.S.; Wouters, P., The leiden ranking 2011/2012: Data collection, indicators, and interpretation. *Journal of the American Society for Information Science and Technology* **2012**, *63*, 2419-2432.

18.    Leydesdorff, L.; Bornmann, L.; Mutz, R.; Opthof, T., Turning the tables in citation analysis one more time: Principles for comparing sets of documents. *Journal of the American Society for Information Science and Technology* **2011**, *62*, 1370-1381.

19.    Bornmann, L.; Leydesdorff, L.; Mutz, R., The use of percentiles and percentile rank classes in the analysis of bibliometric data: Opportunities and limits. *J. Informetr.* **2013**, *7*, 158-165.

20.    Rousseau, R., Basic properties of both percentile rank scores and the i3 indicator. *Journal of the American Society for Information Science and Technology* **2012**, *63*, 416-420.

21.    Schreiber, M., Uncertainties and ambiguities in percentiles and how to avoid them. *Journal of the American Society for Information Science and Technology* **2013**, *64*, 640-643.

22.    Waltman, L.; Schreiber, M., On the calculation of percentile-based bibliometric indicators. *Journal of the American Society for Information Science and Technology* **2013**, *64*, 372-379.

23.    Schubert, T.; Michels, C., Placing articles in the large publisher nations: Is there a "free lunch" in terms of higher impact? *Journal of the American Society for Information Science and Technology* **2013**, n/a-n/a.

24.    Leydesdorff, L.; Radicchi, F.; Bornmann, L.; Castellano, C.; de Nooy, W., Field-normalized impact factors: A comparison of rescaling versus fractionally counted ifs. *Journal of the American Society for Information Science and Technology* **in press**.

25.    Leydesdorff, L.; Bornmann, L., How fractional counting of citations affects the impact factor: Normalization in terms of differences in citation potentials among fields of science. *Journal of the American Society for Information Science and Technology* **2011**, *62*, 217-229.

26.    Radicchi, F.; Fortunato, S.; Castellano, C., Universality of citation distributions: Toward an objective measure of scientific impact. *Proceedings of the National Academy of Sciences* **2008**, *105*, 17268-17272.